\documentclass[twocolumn,superscriptaddress,amsfont,amssymb,amsmath,nofootinbib, showpacs,balancelastpage]{revtex4-1} %[aps,...prl,]
\usepackage{graphicx,longtable,mathrsfs,color}
\usepackage[usenames,dvipsnames]{xcolor} % colour
\usepackage{amssymb,amsmath,mathtools,mathrsfs} % maths
\usepackage{epsfig,subfigure,placeins,float} % plots
\usepackage{booktabs,longtable,ctable} % tables
\usepackage{exscale,relsize} % scale 
\usepackage[normalem]{ulem} % editing
\usepackage{enumerate}

\newcommand{\be}{\begin{equation}}
\newcommand{\ee}{\end{equation}}
\newcommand{\bea}{\begin{eqnarray}}
\newcommand{\eea}{\end{eqnarray}}

\usepackage[stable]{footmisc}

\usepackage{braket}
\usepackage{graphicx}% Include figure files
\usepackage{dcolumn}% Align table columns on decimal point
\usepackage{bm}% bold math
\newcommand{\bml}{\begin{multline}}

\def\dkmu2{\delta K_{\mu \nu}\delta K^{\mu \nu}}
\def\pmu2{  \phi_{\mu \nu}\phi^{\mu \nu}}

\begin{document}

\title{Accelerated expansion in the effective field theory of a radiation dominated universe}

\author{Bruno Balthazar}
\email{bbalthazar@g.harvard.edu}
%\affiliation{Astrophysics,University of Oxford, DWB, Keble Road, Oxford OX1 3RH, UK}
\author{Pedro G. Ferreira}
\email{p.ferreira1@physics.ox.ac.uk}
\affiliation{Astrophysics, University of Oxford, DWB, Keble Road, Oxford OX1 3RH, UK}
\date{Received \today; published -- 00, 0000}

\begin{abstract}
We construct the effective field theory of a perfect fluid in the early universe. Focusing on the case where the fluid has the equation of state of radiation, we  show that it may lead to corrections to the background dynamics that can dominate over those of an effective field theory of gravity alone. We describe the periods of accelerated expansion, in the form of inflationary and bounce solutions, that arise in the background dynamics and discuss their regime of validity within EFT.  
\end{abstract}

%\pacs{98.80.-k, 98.80.Es, 95.36.+d, 95.36.+x}

\maketitle

% PJB: This is temporary, just so we can see what we're doing.
%\tableofcontents

%===============================================================================
%\section{Introduction}
%===============================================================================

An useful tool for modelling physical processes at a well defined energy scale is {\it Effective Field Theory} (EFT) \cite{Weinberg:1978kz}. The idea is to choose an energy scale, $E$, and assume that we know
all the degrees of freedom and associated symmetries at that scale and which are approximately valid up to a higher energy scale, $M$. That new, higher energy scale, marks the onset of new physics - new degrees of freedom or symmetries. One then constructs an action (or a ``phenomenological lagrangian" \cite{Weinberg:1978kz}) which contains
all local operators in the local degrees of freedom, suitably weighted by inverse powers of $M$ \cite{Burgess:1992gx}.
This action should be completely predictive at energies $E$ with errors easily quantifiable \cite{Donoghue:1994dn,Burgess:2003jk,Burgess:2007pt}.

EFT is ideal for understanding some stages of the early Universe. It has been used, with resounding success, to systematise cosmological perturbations during the inflationary regime \cite{Creminelli:2006xe,Cheung:2007st,Weinberg:2008hq,  Matarrese:2007wc, Senatore:2010jy, Bartolo:2010bj, Creminelli:2010qf, Pietroni:2011iz, LopezNacir:2011kk, Achucarro:2012yr, Burgess:2009ea, Barbon:2009ya,Gwyn:2012mw}, and is becoming standard practice in interpreting some aspects of cosmological data \cite{Ade:2013ydc}. It has also been used in other stages of cosmological evolution, such as cosmological acceleration and dark energy models \cite{Park:2010cw,Gubitosi:2012hu, Creminelli:2008wc,Gleyzes:2013ooa, Bloomfield:2011np, Hertzberg:2012qn}, and more recently in modelling large scale structure \cite{Baumann:2010tm, Carrasco:2012cv, Senatore:2014via,Carroll:2013oxa}. 

In this letter we use EFT to understand  the dynamics of the early universe by assuming that we can characterise all the relevant degrees of freedom as a perfect fluid. We use the symmetries of relativistic fluid dynamics to build a self consistent, complete action which we can then study in detail. We find that the fluid will generate some novel behaviour that may be of cosmological significance.
Throughout this paper we will use the convention $\hbar=c=k_B=1$.
%===============================================================================
%\section{The Universe as a perfect fluid.}
%\label{sec: II}

\begin{figure}[h]
\includegraphics[scale=0.46,angle=0]{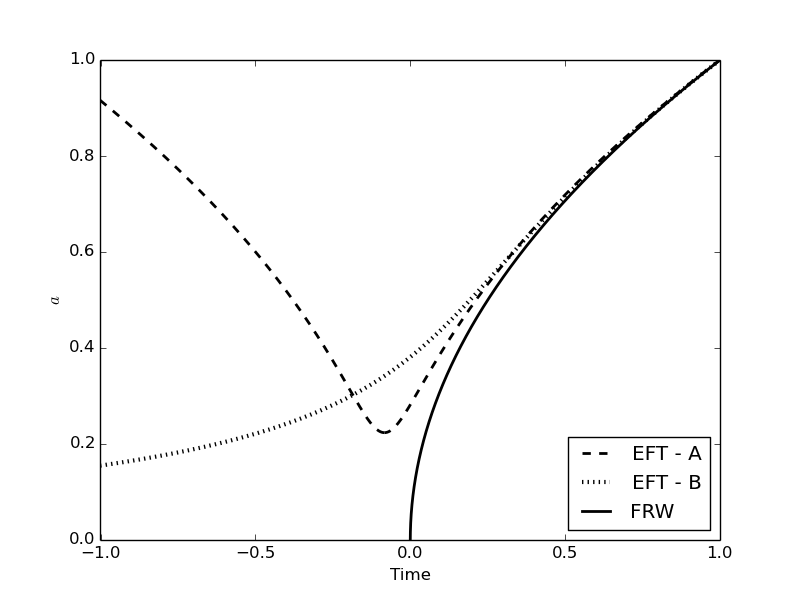}
\centering
\caption{Plot of solutions of equation (\ref{eq: EOM for A}): `EFT - A', equation (\ref{eq: EOM for B}): `EFT - B', and of the classical FRW solution: `FRW', where we have set $M_P=\rho_0=M=1$ and $A=B=0.05$ for clarity (this rescaling does not satisfy $\rho_0\ll M_P^4$ or $M\ll M_P$, however this will not change the behaviour of the solution but only make this new behaviour more evident). The time scale corresponds to one Hubble time ($1/H_0$) in the rescaled units, and the scale for $a$ corresponds to $a(1/H_0)$. Note that at late times all solutions approach the classical solution for a radiation dominated Universe. 
\label{fig: sol for ALL}
}
\end{figure}

At early times, the Universe is a mess of interacting particles and fields.   In principle, the physics of the material content of the early Universe should be complex. In practice, and despite these intrinsic problems, its overall nature is roughly that of a perfect fluid. To understand why this is so, we assume that the Universe is, in general, in thermal equilibrium in the early universe for energies lower (but almost up to) the Planck scale (we know this may not strictly be true above temperatures of $T\simeq 10^{14}$ GeV from QCD alone \cite{Enqvist:1993fm}). Furthermore we assume, that there was no mechanism to inject energy and thermalise the Universe up to that scale.  

We can then define the mean inter particle separation  $\lambda\sim n^{-1/3}$, where $n$ is the number density and the  Hubble scale (or particle horizon) of the Universe is $L\sim H^{-1}$. Consider now the regime where $ {L}/{\lambda}\gg 1$ and take a hyper surface for fixed time $t=t_0$ (which is uniquely defined in the FRW cosmology by the requirements of homogeneity and isotropy). Now choose a volume element with linear scale $D$ such that $\lambda\ll D\ll L$ -- let us call this element a (total) fluid parcel. Inside such an element we may still find many different types of particles, all interacting with each other. Dividing the Universe into fluid parcels we can follow each parcel to a new hyper surface at $t=t_0+\delta t$: from equilibrium, isotropy and homogeneity, we have that the entropy and the average number of each particle species of each fluid parcel remains the same. The same argument follows through, after equilibration, when a new type of particle `joins' the fluid. 

It is clear that we are describing the fluid parcels not based on what type of particles it contains, but by arguing that the fluid will be in equilibrium and furthermore the entropy of each fluid parcel will be conserved. This means that, since all interacting degrees of freedom are present, this fluid is incompressible, as deformations in the fluid volume element could not be dissipated by any other mechanism. As such, the effective theory for a perfect fluid should hold, where we now consider worldlines for the {\it fluid parcels}, and to each we assign a current, such as the entropy current $J^\mu$, as this will be conserved along the worldline. Furthermore, the equation of state for the fluid, $w\equiv P/\rho$ plays a crucial role. 

Naturally, this rationale breaks down near decoupling energies, since the decoupling particles could take away some of the entropy and energy, and hence effectively dissipate it. Furthermore, any injection of energy, such as reheating after an inflationary period, will also affect our arguments. Finally, the fluid approximation will break down at the energy scale where new physics could appear. 

%===============================================================================
%===============================================================================
%\section{Constructing the EFT for a perfect fluid}
%\label{sec:fluid action}

%In this section we discuss what it means to construct an EFT for a perfect fluid in~\ref{sec: what is fluid EFT}, and construct this EFT in \ref{sec: fluid EFT}. We will follow [Dubovsky et al.]
%\subsection{What is the effective field theory of a perfect fluid?}
%The effective Lagrangian includes high energy corrections from the UV completed theory.
%When thinking of an effective field theory, one usually has in mind the low energy dynamics of a scalar field. Of course there is nothing special about a scalar field. What is important is identifying the degrees of freedom in the system.

Consider a perfect fluid in 3+1 dimensions (see \cite{Dubovsky:2005xd, Endlich:2010hf, Dubovsky:2011sj, Nicolis:2011cs, Nicolis:2011ey, Dubovsky:2011sk, Torrieri:2011ne, Hoyos:2012dh, Endlich:2010hf} but also \cite{Ballesteros:2012kv, Ballesteros:2013nwa}). The low-energy degrees of freedom can be chosen to be three scalar fields, $ \phi^I=\phi^I(\vec x,t)$ (where $I=1,2,3$)
which correspond to the comoving (Lagrangian) coordinates of the volume element occupying physical (Eulerian) position $\vec x$ at time $t$. We can choose to align the comoving coordinates with the physical coordinates at a given time $t=t_0$ when the fluid is in equilibrium, so that $\phi^I=x^I$.
With this choice of field variables, the fluid's dynamics must be symmetric under diffeomorphisms $\zeta^I:\phi^I\rightarrow\phi'^I$ that preserve the volume (incompressible fluid), i.e.
$
\phi^I\rightarrow\zeta^I(\phi^J)$
where $\mathrm{det}\left(\frac{\partial\zeta^I}{\partial\phi^J}\right)=1$.
Note that this includes invariance under translations and rotations. In addition, the equations of motion must also be invariant under diffeomorphisms of the space-time coordinates, which means that any tensors appearing in the Lagrangian must be fully contracted.

We now look for the most general relativistic Lagrangian that is invariant under these symmetries. At lowest order, we should have one derivative per field, because of invariance under translations: $
\mathcal{L}_m^{(0)}=\mathcal{L}_m^{(0)}(\partial\phi)$ 
Because of invariance under volume preserving diffeomorphisms, the $\partial\phi^I$ should enter the Lagrangian in the combination
$
J_{\mu\nu\lambda}\equiv\epsilon_{IJK}\partial_{\mu}\phi^I\partial_{\nu}\phi^J\partial_{\lambda}\phi^K
$
where $\epsilon_{IJK}$ is the Levi-Cevita tensor in the matter space, $\epsilon_{123}=1$. This three-tensor does not have a simple physical interpretation, but its dual, $
J^{\mu}\equiv\varepsilon^{\mu\nu\rho\sigma}J_{\nu\rho\sigma}$,
is a vector field along which the conservation laws, 
$
J^{\mu}\partial_\mu\phi^I=0$ and
$\nabla_\mu J^\mu=0$
are both satisfied. Thus it is natural to define the fluid's four-velocity as a unit vector aligned, 
$U^\mu=\frac{1}{b}J^{\mu}$,
where $b\equiv\sqrt{J_\mu J^\mu}$.

From a geometric point of view, $J^\mu$ is the current of fluid points which is aligned with the fluid's four velocity and is conserved. We identify it thus with the entropy current, and $b$ with the entropy density. We will work with the entropy density since like the energy density it should be a conserved quantity even for a mixture of (weakly or strongly) interacting fluids (in the absence of dissipation). Thus, it is the relevant degree of freedom to be used in the EFT construction. 

Consider now
the action  $\mathcal{S}^{(0)}=\int d^4x\sqrt{g}[\frac{M_P^2}{2}R+\mathcal{L}^{(0)}_m(b)]$ where $R$ is the Ricci scalar and $\mathcal{L}_m$ is the fluid Lagrangians. It is then straightforward to find the energy-momentum tensor
\be
T^{(0)}_{\mu\nu}=-b\frac{d\mathcal{L}_m^{(0)}(b)}{db}U_\mu U_\nu-\left(\mathcal{L}_m^{(0)}(b)-b\frac{d\mathcal{L}_m^{(0)}(b)}{db}\right)g_{\mu\nu} \label{perfect}
\ee
This stress-energy tensor has the form of a perfect fluid with density $\rho$, pressure $P$ and equation of state $w$ when
$\rho=-\mathcal{L}_m^{(0)}(b)$, $P=\mathcal{L}_m^{(0)}(b)-b({d\mathcal{L}_m^{(0)}(b)}/{db})$. That is, we have $\mathcal{L}_m^{(0)}=\alpha b^{1+w}$ with $b=\left({-\rho}/{\alpha}\right)^{\frac{1}{1+w}}$ where $\alpha$ is a constant, i.e.  $\mathcal{L}_m(b)=-\rho$.
Note that, since $\mathcal{L}^{(0)}_m(b)$ is invariant under translations, we have$\nabla_\mu T^{\mu\nu}=0$. For the choice $w=1/3$ we have the action for  general relativity in the presence of radiation.

Let us now proceed to construct $\Delta S=\int d^4x\sqrt{g}[\Delta\mathcal{L}(g_{\alpha\beta},b)]$ which consists of higher order terms  involving $J^\mu$, covariant derivatives, and the Riemann tensor (and the related Ricci tensor and scalar). The terms involving only gravity (and no coupling to matter) are found to be \cite{Donoghue:1994dn, Donoghue:2012zc, Donoghue:1993eb}
\be
c_1R^2+c_2R_{\mu\nu}R^{\mu\nu}
\label{eq: eft grav}
\ee
The terms involving only one entropy current $J^\mu$ are
\be
\frac{d_1}{M^2}J^\mu\nabla_\mu R+\frac{d_2}{M^2}\nabla^\mu J^\nu R_{\mu\nu}
\label{eq: eft 1 j}
\ee
since there are no terms in the EFT for the fluid only with a single $J^\mu$ \cite{Dubovsky:2011sj}. The first term can be integrated by parts to zero. 
There are more terms we can write with 2 entropy currents, namely
\bml
\frac{f_1}{M^2}b^2+\frac{e_1}{M^4}J^\mu J^\nu R_{\mu\nu}+\frac{e_2}{M^4}J^\mu J_\mu R+
\\
+\frac{e_3}{M^4}\nabla_\mu J^\nu\nabla_\nu J^\mu+\frac{e_4}{M^4}\nabla_\mu J^\nu\nabla^\mu J_\nu
\label{eq: eft 2 j}
\end{multline}

Of course not all terms have the same suppression power in the cutoff, and so we would expect some of these terms to dominate before others. The reason for keeping all of them however is that they have different origin: the terms in (\ref{eq: eft grav}) come from the EFT for gravity only, those in (\ref{eq: eft 1 j}) are the lowest order in the EFT for matter and gravity, but as we will see later they give a zero contribution to the equations of motion in a homogeneous isotropic Universe, so that in this case it is actually the terms in (\ref{eq: eft 2 j}) that are the lowest order terms in the EFT for gravity and matter. Note that the term with coefficient $f_1$ in (\ref{eq: eft 2 j}) actually comes from the EFT for matter only, which is why we keep the remaining terms with coefficients $e_i$.

Thus, the action for the full effective field theory for gravity with a fluid with an equation of state $w$ is $S=S^{(0)}+\Delta S$ where \begin{widetext}
\bml
\sqrt{g}\Delta\mathcal{L}(g_{\mu\nu},J^\mu)=\sqrt{g}\left[c_1R^2+c_2R_{\mu\nu}R^{\mu\nu}+\frac{d_2}{M^2}\nabla^\mu J^\nu R_{\mu\nu}+\frac{f_1}{M^2}b^2+\frac{e_1}{M^4}J^\mu J^\nu R_{\mu\nu}+\right.
\\
\left.+\frac{e_2}{M^4}J^\mu J_\mu R+\frac{e_3}{M^4}\nabla_\mu J^\nu\nabla_\nu J^\mu+\frac{e_4}{M^4}\nabla_\mu J^\nu\nabla^\mu J_\nu\right].
\label{eq: full Lagrangian}
\end{multline}
\end{widetext}

%\section{Cosmological evolution}
%\label{sec: evol}

Varying $S$ and choosing $w=1/3$ we are now in a position to study the evolution of the Universe governed by the EFT of gravity in the presence of radiation.
For a flat, homogeneous and isotropic Universe we have $g_{\mu\nu}=\mathrm{diag}[1,-a^2(t),-a^2(t),-a^2(t)]
$
and hence the only variable is the scale factor, $a(t)$. We normalise $U^\mu U_\mu=1$ to obtain $J^\mu=\left(b,0,0,0\right)$ in suitable coordinates and find the modified FRW equation
\bml
3M_P^2\left(\frac{\dot a}{a}\right)^2=\rho-A\frac{\rho^\frac{6}{4}}{M^2}-B\frac{\rho^\frac{6}{4}}{M^4}\left(\frac{\dot a}{a}\right)^2+
\\
+C\left[-2\frac{\dddot a}{a}\frac{\dot a}{a}-2\frac{\ddot a}{a}\left(\frac{\dot a}{a}\right)^2+\left(\frac{\ddot a}{a}\right)^2+3\left(\frac{\dot a}{a}\right)^4\right]
\label{eq: full EOM}
\end{multline}
where $A=f_1/\left(-\alpha\right)^\frac{6}{4}$, $B=6(2e_1-5e_2-2e_3-2e_4)/\left(-\alpha\right)\frac{6}{4}$ and $C=12\left(3c_1+c_2\right)$ are all constants, and we have set the contribution of the cosmological constant to zero in the early universe. Note that from the conservation equation $\nabla_\mu J^\mu=0$, and the fact that we chose to start with a radiation fluid, i.e. $\rho\sim b^{\frac{4}{3}}$, leads us to $\rho=\rho_0/a^4$, where $\rho_0$ a constant. Thus, the higher order, fluid, corrections are functions of $\rho$ and the {\it resulting} energy momentum tensor is not that of a perfect  radiation fluid. Alternatively, if one uses Eq. \ref{perfect} to define the new, effective, energy density and pressure of the fluid, one can describe it in terms of a perfect fluid with a time dependent effective equation of state, $w_{\rm eff}(a)\neq1/3$. This point should be emphasised: the $\rho$ in (\ref{eq: full EOM}) is not what one would usually call the energy density of the fluid, after the higher order corrections are included. It is defined as the quantity $-\alpha b^\frac{4}{3}$, which agrees with the energy density of a radiation fluid at lowest order.   %This relation can also be directly checked to be a solution of the energy-conservation equation of the Lagrangian (\ref{eq: full Lagrangian}).

Before we explore the possible solutions it is important to make a few comments on the allowed solutions in an EFT.
The first important thing to note is that, because this is the truncation of an infinite expansion, we are implicitly assuming that the terms in (\ref{eq: full Lagrangian}) dominate the higher order terms. The second important point is that we have a cutoff in the EFT, given by the mass $M$. This means that the solutions we obtain must have characteristic times $\tau\gg1/M$ (see \cite{Simon:1990ic} for a clear discussion of this point). This will usually be related to the first point, i.e. demanding that higher order terms are suppressed. Finally, we will want to recover the classical FRW solution of a radiation dominated Universe at late times.
%\subsection{The space of solutions}
%\label{sec: solutions of EOM}

We now study the effect of each correction term while setting the others to zero. 
\\
\noindent
{\it Case $A\neq0$}: the evolution equation reduces to
\be
3M_P^2\left(\frac{\dot a}{a}\right)^2=\rho-A\frac{\rho^\frac{6}{4}}{M^2}
\label{eq: EOM for A}
\ee
and plot of the solution for $A>0$ is shown in figure \ref{fig: sol for ALL}, while for $A<0$ no interesting new dynamics is found. Defining the conformal time via
$dt=ad\eta$ we can easily find  an analytic solution of the form
$
a=\sqrt{\epsilon\eta^2+a_B^2} \nonumber
$
where $a_B$ and $\epsilon$ are constants.
We can see clearly that the Universe undergoes a bounce, and the minimum value of $a$ is given by $a_B=(A^2\rho_0)^\frac{1}{4}/{M}$ with positive acceleration given by $
\ddot a_B=({M^4}/3A^2M_P^2)a_B
$. 
We also find that $\ddot a=0$ at 
$a_1=\sqrt{2}a_B$ 
at which time the Hubble constant is given by $
H_1^2={M^4}/({24A^2M_P^2})$ 

We must now check if this solution falls in the regime of validity of the EFT we are considering.
Consider the term of the form $\lambda\rho^3/M^8$ with $\lambda$ a constant, i.e. which is a next order term in the EFT for the perfect fluid. At the time of bounce, we have that its relative contribution is of the order ${\lambda}/{A^4}$ which must be much less than $1$. From a similar analysis to other terms, we see that $A\gg1$ is a sufficient condition, if we assume the remaining coefficients such as $\lambda$ to be of order 1.
From our calculation of $H_1$ we have that 
$
\tau\sim A{M_P}/{M}^2
$
but given that $M\ll M_P$ and $A\gg1$ we are sure that this condition is satisfied. From $\ddot{a}_B$ we have $
\tau\sim A^\frac{3}{4}{M_P}/(M^3\rho_0)^\frac{1}{8}
$
which satisfies the conditions required as $\rho_0^\frac{1}{4}\ll M$.
Finally, we have that the correction due to $B\neq0$ will be negligible when $A\gg1$. 

It is interesting to note that, if we include all terms of the form $\rho^\frac{3n}{4}/M^{3n}$ in the perfect fluid EFT, then equation (\ref{eq: EOM for A}) becomes
\be
3M_P^2\left(\frac{\dot a}{a}\right)^2=\rho-f\left(\rho^\frac{3}{4}/M^3\right)\frac{\rho^\frac{6}{4}}{M^2}
\ee
where $f\left(x\right)$ is a Taylor expandable even function in $x$ (it is even because we can only have even powers of $J^\mu$ in the EFT for matter only). This is a more general solution than equation (\ref{eq: EOM for A}), where only the first term in the Taylor expansion is kept. The behaviour of the solution is very similar however, with a point where $\dot a=0$ when $f\left(\rho^\frac{3}{4}/M^3\right)\rho^\frac{1}{2}M^2=1$, which is a bounce for appropriately chosen $f\left(x\right)$, i.e. such that $\ddot a>0$ at this point. We see that the requirement that $A\gg1$ implies that the leading term in the Taylor expansion of $f(x)$ dominates.
\\
\noindent
{\it Case $B\neq0$}: the equation of motion is now given by
\be
3M_P^2\left(\frac{\dot a}{a}\right)^2=\rho-B\frac{\rho^\frac{3}{2}}{M^4}\left(\frac{\dot a}{a}\right)^2
\label{eq: EOM for B}
\ee
and plot of the solution for $B>0$ is shown in figure \ref{fig: sol for ALL}, while for $B<0$ no interesting new dynamics is found. At late times this solution approaches the classical FRW solution, but at early times the terms that dominate the equation of motion (\ref{eq: EOM for B}) are the ones on the right hand side, which means that
$
a(t)\approx-({B^2\rho_0}/{M^8})^\frac{1}{4}({t-t_0})^{-1}
$
at early times, for a constant $t_0$. At the time when $\ddot a=0$, the Universe has a size $a_1=(2B\rho_0^\frac{6}{4}/3M_P^2M^4)^\frac{1}{6}$, and Hubble constant 
$
H_1=(\sqrt{2}M^4/9BM_P)^\frac{1}{3}
$
Note that in order for the EFT still to be valid at this time, we require that $\rho/M^4\ll1$. This means that $B\gg(M_P/M)^2$. Indeed, looking at other terms in the EFT of the form $M^4\nabla^jJ^kR^lM^{-\left(j+2l+3k\right)}$, one finds that all these terms are suppressed at this time if $B$ satisfies this condition.

At late times the solution is clearly the same as the FRW solution, so the only condition left to check is that the solution has a timescale smaller than the cutoff $M$. The timescale $\tau_1$ at the time when $\ddot a=0$ is $\tau_1\sim(BM_P/M^4)^\frac{1}{3}\gg M_P/M^2\gg1/M$ provided that $B\gg(M_P/M)^2$. So we conclude that this condition is necessary and sufficient for this solution to be valid at the time when $\ddot a =0$.
\\
\noindent
{\it Case $C\neq0$}: the equations of motion for this case are well known - they correspond to (extensions of) the well known Starobinsky model of ``$R^2$" inflation \cite{Starobinsky:1980te} and have been extensivley studied in the literature. We won't study this term but we point out that, strictly speaking, if one were to apply the rules of EFT, this correction does not lead to accelerated expansion \cite{Simon:1990ic}; if we were to substitute the classical FRW solution to find the perturbative corrections to the equations of motion, the terms multiplied by $C$ cancel exactly. It is only through the next order corrections that we can get accelerated expansion. 

Nevertheless, this hasn't stopped this model gaining popular support, especially given the recent constraints on the amplitude of tensor modes by the Planck experiment \cite{Ade:2013ydc} (although see also \cite{Ade:2014xna}). In the specific case of $f\left(R\right)$ models of gravity  Ostrogradski's instability can be avoided due to the fact that there is only one second derivative of the metric in $R$ and local constraints on the system arising from gauge invariance. This is not the case of a general EFT however where have to include other contractions of Riemann tensors as well as derivatives of $R$; higher derivatives and other derivatives of the metric crop up, and the constraints on the system are not sufficient to prevent the instability (see \cite{Woodard:2006nt} for more on this). 

%===============================================================================
%\section{Conclusions}
%\label{sec: conclusions}
%===============================================================================
In this paper we have built on the formalism proposed in \cite{Dubovsky:2005xd, Endlich:2010hf, Dubovsky:2011sj, Nicolis:2011cs, Nicolis:2011ey, Dubovsky:2011sk, Torrieri:2011ne, Hoyos:2012dh, Endlich:2010hf} to construct an effective theory of the Universe under the assumption that we can describe the relevant degrees as a perfect fluid. Given that the perfect fluid assumption is at the heart of our current model of the universe, and specifically for the dynamics of the scale factor, we believe it is a conservative one. 

We have found that, starting with a radiation fluid, the resulting energy momentum tensor is not that of a radiation fluid. More importantly, amongst our solutions, we find accelerated expansion at early times. This is remarkable because we have not explicitly added any field to do the job for us - this behaviour arises naturally in the most conservative model of the early universe, an FRW space-time with radiation. It is the (expected) corrections that arise in EFT that lead us to these solution, i.e. there is no new physics invoked. To be more specific, the interesting regimes correspond to taking either $A$ or $B$ as non-zero - the dynamics for each of these cases is illustrated in Figure \ref{fig: sol for ALL}. 

In both cases we have found that we are, strictly, breaking the rules of EFT. In case A we have found that $A\gg1$, where we would ideally have $A\sim 1$.
 In case B, the situation is much worse: to have effect, we need $B\gg\left(M_P/M\right)^2$, which is well beyond what is acceptable. Nevertheless, this may not be a show stopper as can be seen in the case of Higgs inflation \cite{Shaposhnikov:2009pv} or $R^2$  inflation \cite{Starobinsky:1980te}; these theories suffer from the same problem yet they continue to be considered seriously in a number of studies. 

In case A  we found a smooth bounce occurring well within the regime of validity of the EFT, Consistent bounce solutions are rare and far between and can be incredibly useful in understanding the origin and evolution of perturbations arising in a pre-Big Bang era \cite{Allen:2004vz}. In case B we found a bona-fide inflating solution- interestingly enough this arises in the context of a minimal extension to the standard model, along the lines of \cite{Starobinsky:1980te}. For both of these cases, and given that we have laid out the complete framework for how to build the EFT for our Universe, the next obvious step is to work out the origin and evolution of perturbations. This is of particular relevance for the bouncing model where there is a dearth of consistent models but, more importantly, these calculations can then be compared to the hugely successful calculations arising in the (by now) standard formalism for the EFT of inflation as promoted in \cite{Cheung:2007st,Weinberg:2008hq}. 

\textit{Acknowledgments.---} We thank T.Baker, J.Bonifacio, C. Burgess, M. Lagos, J.March-Russell, A. Maroto, J. Noller, G.Ross, J. Scargill and L.Senatore for useful discussions. PGF acknowledges support from Leverhulme, STFC, BIPAC and the Oxford Martin School.

%===============================================================================

%===============================================================================
% BIBLIOGRAPHY
%===============================================================================

\bibliographystyle{apsrev4-1}
\bibliography{FluidEFT}

\end{document}